\documentclass[a4paper]{jpconf}
\usepackage{graphicx}
 \usepackage{amsmath}
\begin{document}
\title{New physics in top decay}

\author{Celine Degrande}

\address{Department of Physics, University of Illinois at Urbana-Champaign\\1110 W. Green Street, Urbana, IL 61801}

\ead{cdegrand@illinois.edu}

\begin{abstract}
After a short introduction to effective field theories, most of their features are illustrated using the top decay. The  effects of heavy new physics on the top decay are computed and the constraints on the coefficients of the dimension-six operators are derived from the available measurements.  
\end{abstract}

\section{Introduction}

The outstanding performances of the LHC have led last summer to the discovery of a new particle compatible with the Higgs boson. However, the LHC has not been built only to find the last missing piece of the Standard Model (SM) but also to find new physics. If the new degrees of freedom are light enough to be produced by the LHC, the new physics would manifest itself as new resonances. On the contrary, new physics would appear as anomalous interactions between the known particles if the center of mass energy is lower than the scale of new physics. 
This paper focus on the second case. The theoretical framework will be introduced in section \ref{eft} and will be applied to top decay in section \ref{topdecay}.

\section{Effective field theories}\label{eft}

Effective field theories (EFT) rely on the hierarchy between the energy reached in the experiment and the scale of new physics. Since their  ratio can only be used as an expansion parameter if it is smaller than one, an EFT is only valid below the new physics scale, $\Lambda$. 
In the Lagrangian, the new interactions arise from higher dimensional operators suppressed by the new physics scale,\begin{equation}
 \mathcal{L} = \mathcal{L}_{SM} + \sum_{d>4}\sum_{i} \frac{c_i^d}{\Lambda^{d-4}} \mathcal{O}_i^d,
\end{equation} 
where $d$ is the dimension of the operator $ \mathcal{O}_i^d$ and $c_i^d$ are coefficients which can be computed from the high energy theory.
Far below the new physics scale, only the operators with the lowest dimension are required for a given precision. Their coefficients can be kept as free parameters to be model independent since the number of relevant operators is finite.
However, the infinite sum of operators is needed as the new scale is reached by the experiment and the effective theory loses its predictive power. \\
The new physics at the Tevatron or at the LHC can only be described by an EFT if the associated scale is at or above one TeV. Therefore the low energy degrees of freedom are the SM fields including the Higgs field. We assume also that the SM gauge symmetries as well as the Baryon and Lepton numbers are preserved in agreement with the experimental data. 
Those requirements imply that only operators with an even dimension can be built \cite{Degrande:2012wf}. Consequently, the largest new physics contributions are  due to the dimension-six operators. 
While there are already 59 dimension-six operators for one generation of fermions~\cite{Buchmuller:1985jz,Grzadkowski:2010es}, only a few can affect a specific process~\cite{Buchmuller:1985jz,Leung:1984ni} and they can often be distinguished by their different effects. 
Due to the symmetries and the scale suppression, the model is more predictive than the anomalous couplings approach and therefore provides some guidance in the quest for new physics. In fact, anomalous couplings Lagrangians  only satisfy Lorentz invariance and usually electromagnetic gauge invariance is imposed afterwards while the $SU(2)_L$ is ignored. Only the operators with the lowest number of derivatives are kept without any justification as the scale suppression is lacking.  
Moreover,  the effective Lagrangian can be used for any process because gauge invariance is guaranteed by construction. Like in the SM, gauge invariance requires that vertices with different numbers of gauge bosons are related to each other. On the contrary, all the vertices from an anomalous couplings Lagrangian are independent. 
Since the effective field theory is renormalizable in the modern sense, it can be used in loop computation. Therefore the parameters can be constrained both by direct and indirect measumements. 
Finally, unitarity is satisfied in the EFT validity region, \textit{i.e.} below the new scale and no form factors are needed \cite{Degrande:2012wf}. However, this last point can hardly be illustrated for top decay as the energies of the decay products are bound by the top mass.

\section{Top decay}\label{topdecay}

We will focus here on the SM-like top decay, namely $t\to b W$ with the W decaying eventually in a pair of leptons or light quarks. Exotic decay channels can be found for example in Ref.~\cite{AguilarSaavedra:2010zi}.

\subsection{The operators}

\begin{figure}[h]
\includegraphics[width=0.6\textwidth]{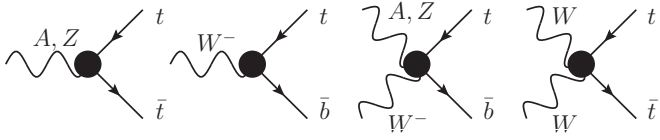}\hspace{0.05\textwidth}%
\begin{minipage}[b]{0.35\textwidth}\caption{\label{fig:otwvert}Vertices of the $\mathcal{O}_{tW} $ operators. For each vertex in this figure, there is one extra vertex with an additional Higgs line.}
\end{minipage}
\end{figure}

In the massless limit for the b quark\footnote{Extra operators can contribute to the $t \to Wb$  decay if this assumption is removed \cite{delAguila:2000aa}.}, only two operators affect the $t \to b W $ decay at the order $\Lambda^{-2}$\cite{Zhang:2010px},
\begin{equation}
\mathcal{O}_{\phi q}^{(3)} = i\left(\phi^\dagger \tau^i D_\mu \phi\right) \left(\bar{q}\gamma^\mu \tau^i q\right)+h.c.\qquad\text{and}\qquad
\mathcal{O}_{tW} = \bar{q} \sigma_{\mu\nu} \tau^i  t \tilde{\phi} W^{\mu\nu}_i. \label{eq:optbw}
\end{equation}
Keeping only the terms with the top, the bottom and one W field and comparing to the anomalous Lagrangian \cite{Kane:1991bg},
\begin{equation}
\mathcal{L}_{AC} = \frac{g}{\sqrt2} \bar{b} \gamma^\mu V_{tb} \left(f_V^L\gamma_L + f_V^R\gamma_R\right) t W_\mu - \frac{ig}{\sqrt2 m_W} \bar{b}  \sigma^{\mu\nu} q_\nu V_{tb} \left(f_V^L\gamma_L + f_V^R\gamma_R\right)t W_\mu +h.c.\label{eq:anoL},
\end{equation}
the prediction of the effective field theory for the anomalous couplings reads
\begin{equation}
\begin{array}{lll}
f_V^L = \frac{C_{\phi q}^{(3)} v^2}{V_{tb}\Lambda^2} &\quad &f_V^R=0\\
f_T^L =0 && f_T^R=-\sqrt2 \frac{ C_{tW}  v^2}{ V_{tb}\Lambda^2}.\\
\end{array}
\end{equation}
As already mentioned, the anonalous Lagrangian in Eq.~\eqref{eq:anoL} is not completely generic as further terms with extra  momenta can be added.
The effective field theory is more predictive since it has two parameters less than anomalous couplings for the $Wtb$ vertex but also because it fixes some of the anomalous couplings of other vertices.  For example, the vertices generated by $\mathcal{O}_{tW} $ are shown on Fig.~\ref{fig:otwvert}. Those vertices are required in higher multiplicity processes like $t \to bW\gamma$ to insure gauge invariance. If only the anomalous $tbW$ vertex  from Eq.~\eqref{eq:anoL} was added to the SM Lagrangian for this process, the result would not be gauge invariant.

In addition to the operators in Eq.~\eqref{eq:optbw}, one operator contribute to the $t\to b l\bar\nu_l $ decay\cite{AguilarSaavedra:2010zi} without the exchange of a W boson,
\begin{eqnarray}
\mathcal{O}_{ql}^{(3)} &=& \left(\bar{q}\gamma^\mu \tau^i q\right) \left(\bar{l}\gamma_\mu \tau^i l\right)\label{eq:optbln}
\end{eqnarray}
and a similar operator contribute to the $t\to b d\bar u $ decay. With all those dimension-six operators, the square matrix element for $t\to b l\bar\nu_l $ reads
\begin{eqnarray}
\frac{1}{2}\Sigma|M|^2&=&\frac{V_{tb}^2g^4u(m_t^2-u)}{2(s-m_W^2)^2}\left(1+2\frac{C_{\phi q}^{(3)}v^2}{V_{tb}\Lambda^2}\right)+\frac{4\sqrt{2}{\rm Re}C_{tW}V_{tb}m_tm_W}{\Lambda^2}\frac{g^2su}{(s-m_W^2)^2}\nonumber\\
&&+\frac{4C_{ql}^{(3)}}{\Lambda^2}\frac{g^2u(m_t^2-u)}{s-m_W^2}+ \mathcal{O}\left(\Lambda^{-4}\right)\label{eq:m2}
\end{eqnarray}
where $s\equiv(p_t-p_b)^2$ and $u\equiv(p_t-p_e)^2$.
The contribution of $\mathcal{O}_{\phi q}^{(3)}$ is proportional to the SM one because they generate the same $Wtb$ vertex up to a global factor. 

\subsection{The width}

The width is affected mainly by $\mathcal{O}_{\phi q}^{(3)}$ and $\mathcal{O}_{tW}$, 
\begin{equation}
\frac{\Gamma\left(t \to b e^+ \nu_e\right)}{GeV} = 0.1541 + \left[0.019 \frac{C_{\phi q}^{(3)}}{\Lambda^2} 
+ 0.026 \frac{C_{tW}}{\Lambda^2} + 0 \frac{C_{ql}^{(3)}}{\Lambda^2}\right]\text{TeV}^2 \label{eq:width}.
\end{equation}
The contribution of the four-fermion operator almost vanishes. The measured value \cite{Abazov:2012vd} together with the SM prediction \cite{Jezabek:1988iv} give a first constraint on the coefficients of the two relevant operators,
\begin{equation}
 \left.
 \begin{array}{r}
  \Gamma_{meas} = 2^{+0.47}_{-0.43}\,\text{GeV}\\[3mm]
  \Gamma_{SM} = 1.33\,\text{GeV}
 \end{array}
 \right\} \frac{C_{\phi q}^{(3)}}{\Lambda^2} 
 + 1.35 \frac{C_{tW}}{\Lambda^2} = 4^{+2.8}_{-2.5}\text{TeV}^{-2}.\label{eq:widthconst}
  \end{equation}

\subsection{The W helicities}

The helicity of the W boson affects the angular distribution of top decay products defined by the helicity fractions,
\begin{equation}
\frac{1}{\Gamma}\frac{d\Gamma}{d\cos\theta}\equiv \frac{3}{8}(1+\cos\theta)^2F_R +\frac{3}{8}(1-\cos\theta)^2F_L +\frac{3}{4}\sin^2\theta F_0 
\end{equation}
where $\theta$ is the angle between the top and the neutrino momenta in the W rest frame.
The helicity fractions are only affected by $\mathcal{O}_{tW}$,
\begin{eqnarray}
F_0&=&\frac{m_t^2}{m_t^2+2m_W^2}-\frac{4\sqrt{2}{\rm Re}C_{tW}v^2}{\Lambda^2V_{tb}}\frac{m_tm_W(m_t^2-m_W^2)}{(m_t^2+2m_W^2)^2}\nonumber\\
F_L&=&\frac{2m_W^2}{m_t^2+2m_W^2}+\frac{4\sqrt{2}{\rm Re}C_{tW}v^2}{\Lambda^2V_{tb}}\frac{m_tm_W(m_t^2-m_W^2)}{(m_t^2+2m_W^2)^2}\nonumber\\
F_R&=&0.
\end{eqnarray}
$F_R$ vanishes because the b quark is massless and the sum of the helicity fractions is one by definition. The SM prediction at NNLO \cite{Czarnecki:2010gb} and the Atlas result  \cite{Aad:2012ky} with the constraints $F_0+F_L+F_R=1$ and $F_R=0$  strongly limit the coefficient of  $\mathcal{O}_{tW}$
\begin{equation}
\left.\begin{array}{r}
 F_0 ^{SM}= 0.687\pm5\\[3mm]
  F_0^{meas} = 0.66\pm5
 \end{array}\right\}
 \frac{C_{tW}}{\Lambda^2} = 0.44\pm0.9 \text{TeV}^{-2} .\label{eq:ctwconst}
\end{equation}
Together with the constraint from the total width (\ref{eq:widthconst}), this result fixes the range allowed for the coefficient of $\mathcal{O}_{\phi q}^{(3)}$
\begin{equation}
 \frac{C^{(3)}_{\phi q}}{\Lambda^2} = 3.4^{+4}_{-3.7} \text{TeV}^{-2} .
\end{equation}

\subsection{The leptons invariant mass}

\begin{figure}[h]
\includegraphics[width=0.6\textwidth]{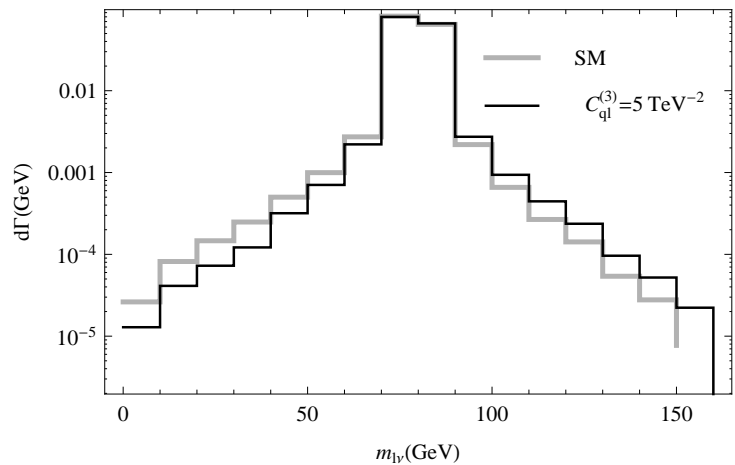}\hspace{0.05\textwidth}%
\begin{minipage}[b]{0.35\textwidth}\caption{\label{fig:mlnu}Invariant mass of the leptons pair in the top decay $t \to be\bar\nu_e$ for the SM and the SM and $\mathcal{O}_{ql}^{(3)}$ at the order $\Lambda^{-2}$, \textit{i.e.} only the interference between the SM and  the four-fermion operator is added to the SM distribution.}
\end{minipage}
\end{figure}

As shown in Eq.~(\ref{eq:m2}), the contribution of the four-fermion operator only differs from the SM by the power of the denominator. In fact, the remaining $\left(s-m_W^2\right)$ factor is due to the SM amplitude. Consequently, $\mathcal{O}_{ql}^{(3)}$ contribution on the leptons invariant mass changes of sign at the W mass as displayed on Fig.~\ref{fig:mlnu}. 
Accidentally, the contributions from the two regions are almost equal in modulus such that their sum nearly vanishes. 
The effect of the corresponding four-fermion operator for the light quarks is identical. The four-fermion operator cannot be constrained yet because the leptons invariant mass has not been measured so far.

\subsection{An example of indirect constraints}

\begin{figure}[b]
\includegraphics[width=0.6\textwidth]{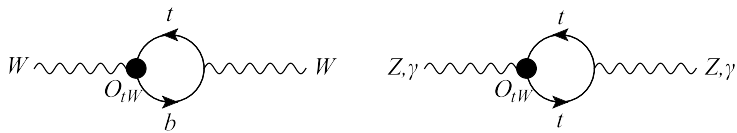}\hspace{0.05\textwidth}%
\begin{minipage}[b]{0.35\textwidth}\caption{\label{fig:loop}One loop corrections to the two points functions of the electroweak bosons due to the operator  $\mathcal{O}_{tW}$.}
\end{minipage}
\end{figure}

A modification of the $Wtb$ vertex affects the two points function of the W boson at the loop level as shown on the left diagram in Fig.~\ref{fig:loop}. 
However, those corrections have infinities and are useless unless they can be absorbed using renormalization.  Contrary to anomalous couplings, an EFT is renormalizable order by order in $\Lambda$.  The contribution from $\mathcal{O}_{tW}$ to the $\widehat{S}$ parameter is infinite and absorbed by the renormalization of the coefficient of a dimension-six operator affecting $\widehat{S}$ at the tree-level while the contribution to $\widehat{T}$ vanishes. However, the correction to the $\widehat{U}$ parameter is finite \cite{Greiner:2011tt} and can be used to constrain the coefficient of this operator,
\begin{equation}
\left.
\begin{array}{r}
 \widehat{U} = N_c\frac{gC_{tW}}{4\pi^2}\frac{\sqrt{2}vm_t}{4\Lambda^2}\\[3mm]
 \widehat U_{meas}=(-5.0 \pm 8.4)\times 10^{-4}
\end{array}
\right\}\frac{C_{tW}}{\Lambda^2}=-0.7\pm1.1\ {\rm TeV}^{-2}.
\end{equation}
This constraint is comparable and in agreement with the one derived from the W helicities and presented in Eq.~\eqref{eq:ctwconst}. Similar indirect constraints can be obtained from other processes as well. For example, the $Wtb$ vertex also alters $b\to s \gamma$~\cite{Martinez:1994my}.

\section{Conclusion}
 
If the new physics is heavy, it should affect the top width, the W helicity fractions and the invariant mass distribution of the leptons or light quarks according to the effective extension of the SM. All the four dimension-six operators relevant for the top decay are or can be constrained by measuring those quantities. However, the top width is poorly known and the invariant mass distribution has no been measured yet. Those operators can also be constrained using indirect measurements because effective field theories are renormalizable in the moderne sense.
We have shown in addition that the effective field approach is more predictive and simpler, i.e. guarantee gauge invariance, than the alternative anomalous couplings approach.
Finally, EFT can now be easily implemented into MadGraph thanks to the new FeynRules interface~\cite{Degrande:2011ua} and ALOHA~\cite{deAquino:2011ub}. In particular, the effective theory for top decay is available in MadGraph5~\cite{TopEffTh}.

\section*{References}
\bibliography{biblio}

\providecommand{\newblock}{}
\begin{thebibliography}{10}
\expandafter\ifx\csname url\endcsname\relax
  \def\url#1{{\tt #1}}\fi
\expandafter\ifx\csname urlprefix\endcsname\relax\def\urlprefix{URL }\fi
\providecommand{\eprint}[2][]{\url{#2}}

\bibitem{Degrande:2012wf}
Degrande C, Greiner N, Kilian W, Mattelaer O, Mebane H {\em et~al.\/} 2012
  (\textit{Preprint} \eprint{1205.4231})

\bibitem{Buchmuller:1985jz}
Buchmuller W and Wyler D 1986 {\em Nucl.Phys.\/} {\bf B268} 621

\bibitem{Grzadkowski:2010es}
Grzadkowski B, Iskrzynski M, Misiak M and Rosiek J 2010 {\em JHEP\/} {\bf 1010}
  085 and ref. [10-13] therein (\textit{Preprint} \eprint{1008.4884})

\bibitem{Leung:1984ni}
Leung C~N, Love S and Rao S 1986 {\em Z.Phys.\/} {\bf C31} 433

\bibitem{AguilarSaavedra:2010zi}
Aguilar-Saavedra J 2011 {\em Nucl.Phys.\/} {\bf B843} 638--672
  (\textit{Preprint} \eprint{1008.3562})

\bibitem{delAguila:2000aa}
del Aguila F, Perez-Victoria M and Santiago J 2000 {\em Phys.Lett.\/} {\bf
  B492} 98--106 (\textit{Preprint} \eprint{hep-ph/0007160})

\bibitem{Zhang:2010px}
Zhang C and Willenbrock S 2010 {\em Nuovo Cim.\/} {\bf C033N4} 285--291
  (\textit{Preprint} \eprint{1008.3155})

\bibitem{Kane:1991bg}
Kane G~L, Ladinsky G and Yuan C 1992 {\em Phys.Rev.\/} {\bf D45} 124--141

\bibitem{Abazov:2012vd}
Abazov V~M {\em et~al.\/} (D0 Collaboration) 2012 {\em Phys.Rev.\/} {\bf D85}
  091104 (\textit{Preprint} \eprint{1201.4156})

\bibitem{Jezabek:1988iv}
Jezabek M and Kuhn J~H 1989 {\em Nucl.Phys.\/} {\bf B314} 1

\bibitem{Czarnecki:2010gb}
Czarnecki A, Korner J~G and Piclum J~H 2010 {\em Phys.Rev.\/} {\bf D81} 111503
  (\textit{Preprint} \eprint{1005.2625})

\bibitem{Aad:2012ky}
Aad G {\em et~al.\/} (ATLAS Collaboration) 2012 {\em JHEP\/} {\bf 1206} 088
  (\textit{Preprint} \eprint{1205.2484})

\bibitem{Greiner:2011tt}
Greiner N, Willenbrock S and Zhang C 2011 {\em Phys.Lett.\/} {\bf B704}
  218--222 (\textit{Preprint} \eprint{1104.3122})

\bibitem{Martinez:1994my}
Martinez R, Perez M and Toscano J 1994 {\em Phys.Lett.\/} {\bf B340} 91--95

\bibitem{Degrande:2011ua}
Degrande C, Duhr C, Fuks B, Grellscheid D, Mattelaer O {\em et~al.\/} 2012 {\em
  Comput.Phys.Commun.\/} {\bf 183} 1201--1214 (\textit{Preprint}
  \eprint{1108.2040})

\bibitem{deAquino:2011ub}
de~Aquino P, Link W, Maltoni F, Mattelaer O and Stelzer T 2012 {\em
  Comput.Phys.Commun.\/} {\bf 183} 2254--2263 (\textit{Preprint}
  \eprint{1108.2041})

\bibitem{TopEffTh}
\urlprefix\url{https://cp3.irmp.ucl.ac.be/projects/madgraph/wiki/Models/TopEff%
Th}

\end{thebibliography}

\end{document}